\documentclass[aoas,nameyear,MSNbibl,dvips]{arximspdf}
\usepackage{graphicx}

\doi{10.1214/10-AOAS398B}
\referstodoi{10.1214/10-AOAS398}
\volume{5}
\issue{1}
\pubyear{2011}
\firstpage{76}
\lastpage{79}

\begin{document}
\begin{frontmatter}

\title{Discussion of: A statistical analysis of multiple temperature
proxies: Are
reconstructions of surface temperatures over the last 1000~years~reliable?\thanksref{T1,T2}}
\runtitle{Discussion}
\pdftitle{Discussion on A statistical analysis of multiple temperature
proxies: Are reconstructions of surface temperatures over the last 1000
years reliable?
by B. B. McShane and A. J. Wyner}

\begin{aug}
\author{\fnms{Jason E.} \snm{Smerdon}\corref{}\ead[label=e1]{jsmerdon@ldeo.columbia.edu}%
\ead[label=u1,url]{http://www.ldeo.columbia.edu/\textasciitilde jsmerdon}}

\runauthor{J. E. Smerdon}

\affiliation{Lamont-Doherty Earth Observatory of Columbia University}

\address{Lamont-Doherty Earth Observatory\\
61 Route 9W \\
P.O. Box 1000\\
Palisades, New York 10964\\
USA\\
\printead{e1}\\
\printead{u1}} 
\end{aug}
\thankstext{T1}{Lamont-Doherty Earth Observatory contribution number 7439.}
\thankstext{T2}{Supported in part by NSF Grant ATM-09-02436 and NOAA
Grant NA07OAR4310060.}

\received{\smonth{9} \syear{2010}}


%

\end{frontmatter}

McShane and Wyner [(\citeyear{McShane2010}); hereinafter MW11] reiterate a well-known and
central challenge of paleoclimatology: it is fraught with uncertainties
and based on noisy observations. Decades of research have aimed at
characterizing these uncertainties and interpreting proxies through
laboratory experiments, field observations, theory, process-based
modeling, cross-record comparisons, and indeed through statistical
modeling and hypothesis testing. It is against this larger backdrop
that the problem addressed by MW11 must be considered. Attempts to
reconstruct global or hemispheric temperature indices and fields using
multi-proxy networks are an outgrowth of many efforts in
paleoclimatology, but represent relatively recent pursuits in the
field. They provide neither the principal scientific evidence
supporting climate-proxy connections, nor the most compelling, and the
inference by MW11 that their own findings demonstrate a widespread
failure in the predictive capacity of climate proxies is at odds with
most other independent lines of proxy research.

The above considerations notwithstanding, I focus on one principal
argument by MW11 that uses cross-validation experiments to conclude
that ``\textit{proxies are severely limited in their ability to predict
average temperatures and temperature gradients.}'' I demonstrate that
this claim is based on a hypothesis test subject to Type~II errors and
therefore an inconclusive evaluation of the temperature sensitivity of
proxy archives.

\begin{figure}

\includegraphics{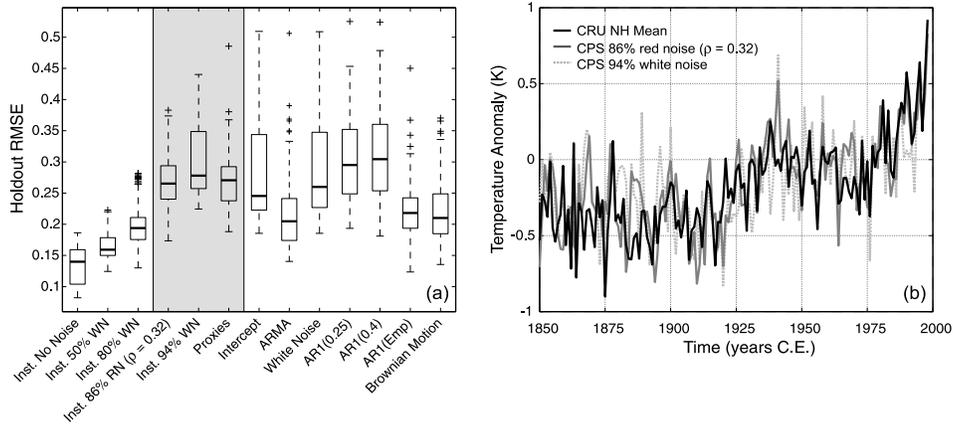}

\caption{\textup{(a)} Cross-validated RMSE on $30$-year hold out blocks
for a subset of the original MW11 experiments (recomputed by the
current author) and for the newly completed experiments using the
instrumental data perturbed with various levels of noise. The $86$\%
red-noise experiment is considered to be representative of the average
proxy predictor [Mann et al. (\protect\citeyear{Mann2007})] and compares well to the proxy
result. \textup{(b)}  Area-weighted CPS reconstructions using the $86$\%
red-noise and $94$\% white-noise predictor sets. The means of the
reconstructed time series have been set equal to the mean of the CRU
target series during the 1961--1990 C.E. interval.}\label{fig1}
\end{figure}

I perform additional cross-validation experiments using 283 time series
that are randomly selected from the global CRU temperature field as
infilled and subselected (1732 total grid cells) by Mann et al.
(\citeyear{Mann2008}). I choose 283 samples based on the total number of unique
$5^{\circ} \times5^{\circ}$ grid cells that contain the 1209 proxies
in the Mann et al. (\citeyear{Mann2008}) proxy network. Time series from these cells
are used to create five predictor datasets spanning the instrumental
period by adding 0, 50, 80 and 94\% white noise by variance, and 86\%
red noise by variance ($\rho= 0.32$), the latter of which has been
argued to be an average representation of the noise in proxy records
[Mann et al. (\citeyear{Mann2007})]. These datasets are used to repeat the MW11
analysis using the Lasso to target the CRU NH mean index across 120
cross-validation experiments. My results are summarized in Figure~\ref{fig1}(a)
and compared to my own reproduction of some MW11 experiments [see
Smerdon (\citeyear{Smerdon2010}) for supplementary code and data].

I first note that even the no-noise experiment is subject to errors.
This is an important baseline for the MW11 experiments, demonstrating
that even ``perfect proxies'' are subject to errors due to incomplete
field sampling. Most importantly, however, the skill of the
instrumental predictors diminishes with noise such that the 86\%
red-noise and 94\% white-noise predictors perform \textit{comparable to or
worse than} the proxy network, and in turn perform worse than the
AR1(Emp) and Brownian motion null models tested by MW11. Additionally,
simple area-weighted composite-plus-scale (CPS) ``reconstructions''
from the Northern Hemisphere (NH) subset of the 86\% red-noise and 94\%
white-noise predictor networks yield NH mean indices that compare well
with the target [Figure~\ref{fig1}(b); respective correlations between the target
and CPS reconstructions are 0.73 and 0.62], indicating that skillful
reconstructions are possible from networks with such noise levels.
These findings are fundamental to the MW11 argument that the proxies
are poor temperature recorders because they do not perform better than
some noise models. To the contrary, the results that I present
demonstrate that predictor networks explicitly containing temperature
signals---perturbed with approximate proxy noise levels---also do not
beat the AR1(Emp) and Brownian motion noise models in cross-validation
experiments and that skillful CPS reconstructions can be derived from
such predictors. The appropriate conclusion is therefore not that the
proxies are limited in their ability to predict NH temperatures, but
that the test performed by MW11 is subject to Type~II errors and is
unsuitable for measuring the degree to which the proxies sample
temperature. Note that this conclusion, although challenged by MW11,
also supports arguments by Ammann and Wahl (\citeyear{Ammann2007}) about the dangers of
Type II errors in this paleoclimatic context.

It is worth considering the likely reasons why the AR1(Emp) and
Brownian motion noise models perform better than predictors explicitly
containing temperature signals. As discussed by MW11, the large amounts
of persistence in these models are good approximations of a principal
characteristic of the target time series, namely its temporal
autocorrelation. This fact, combined with the short cross-validation
period, allows highly persistent time series to test well. But this
success is likely also dependent on selections from many noise draws.
MW11 have focused on the Mann et al. (\citeyear{Mann2008}) study that includes 1209
total proxies (1138 if the Lutannt series are excluded) and thus a
large number of possible predictors in the MW11 noise experiments. In
contrast, other NH temperature reconstructions have been successfully
cross validated using only a few tens of proxies [e.g., Esper, Cook and Schweingruber
(\citeyear{Esper2002}); Moberg et al. (\citeyear{Moberg2005}); Hegerl et al. (\citeyear{Hegerl2007})]. It therefore is
yet unclear how the MW11 cross-validation tests would compare in
scenarios using far fewer predictors.

MW11 present a number of experiments that deserve further testing and
analyses. The issues that I raise similarly come with their own set of
caveats that cannot be explored in a short discussion paper. For
example, a field sampling reflecting the true Mann et al. (\citeyear{Mann2008}) proxy
locations with reduced ocean sampling and regional clustering might
worsen the cross-validation skill of the instrumental predictors
compared to the random sampling used in my experiments. Conversely, the
true proxy distribution is more concentrated in the NH, which may
improve prediction of the NH mean index. I have also sampled each grid
cell once, as opposed to multiple sampling reflecting the occurrence of
several proxies in a single grid cell. This latter sampling will in
effect reduce the noise in the relevant cells, thus making the noise
dependence of cross-validation skill less straightforward to interpret.
These dependencies should be tested in future work. Nevertheless, the
preliminary results that I have outlined suggest that the MW11
hypothesis test is subject to Type~II errors and thus is not suitable
for evaluating the reliability of proxy archives as temperature predictors.

\section*{Acknowledgments}
I thank Alexey Kaplan for many insightful discussions on the subject of
this manuscript and Blakeley McShane for helpful technical exchanges.

\begin{supplement}
\stitle{Code and data files}
\slink[doi]{10.1214/10-AOAS398BSUPP} 
\slink[url]{http://lib.stat.cmu.edu/aoas/398B/supplementB.zip}
\sdatatype{.zip}
\sdescription{This supplement comprises a zip file containing the code
and data files used for this
manuscript and a README document describing all the files included in
the directory.}
\end{supplement}

%

\printaddresses

\end{document}